\documentclass[aps,prl,preprint,showpacs,superscriptaddress,amsmath,amssymb]{revtex4}
\usepackage[english]{babel} 
\usepackage{color} 
\usepackage{graphicx}
\usepackage{amsmath}
\usepackage{amsfonts}
\usepackage{amssymb}
\usepackage[squaren]{SIunits}

\newcommand{\tav}[1]{\langle #1 \rangle} 
\newcommand{\tcm}[2]{\tav{( #1-\tav{#1})^#2}} 

\begin{document}

\title{Dual-path state reconstruction scheme for propagating quantum microwaves and detector noise tomography}

\author{E.~P.~Menzel}
\email{Edwin.Menzel@wmi.badw-muenchen.de}
\affiliation{Walther-Mei{\ss}ner-Institut, Bayerische Akademie der Wissenschaften, D-85748~Garching, Germany}
\affiliation{Physik-Department, Technische Universit\"{a}t M\"{u}nchen, D-85748 Garching, Germany}

\author{F.~Deppe}
\affiliation{Walther-Mei{\ss}ner-Institut, Bayerische Akademie der Wissenschaften, D-85748~Garching, Germany}
\affiliation{Physik-Department, Technische Universit\"{a}t M\"{u}nchen, D-85748 Garching, Germany}

\author{M.~Mariantoni}
\altaffiliation[Present address: ]{Department of Physics, University of California, Santa Barbara, California 93106, USA}
\affiliation{Walther-Mei{\ss}ner-Institut, Bayerische Akademie der Wissenschaften, D-85748~Garching, Germany}
\affiliation{Physik-Department, Technische Universit\"{a}t M\"{u}nchen, D-85748 Garching, Germany}

\author{M.~\'{A}.~\surname{Araque~Caballero}}
\affiliation{Walther-Mei{\ss}ner-Institut, Bayerische Akademie der Wissenschaften, D-85748~Garching, Germany}
\affiliation{Physik-Department, Technische Universit\"{a}t M\"{u}nchen, D-85748 Garching, Germany}

\author{A.~Baust}
\affiliation{Walther-Mei{\ss}ner-Institut, Bayerische Akademie der Wissenschaften, D-85748~Garching, Germany}
\affiliation{Physik-Department, Technische Universit\"{a}t M\"{u}nchen, D-85748 Garching, Germany}

\author{T.~Niemczyk}
\affiliation{Walther-Mei{\ss}ner-Institut, Bayerische Akademie der Wissenschaften, D-85748~Garching, Germany}
\affiliation{Physik-Department, Technische Universit\"{a}t M\"{u}nchen, D-85748 Garching, Germany}

\author{E.~Hoffmann}
\affiliation{Walther-Mei{\ss}ner-Institut, Bayerische Akademie der Wissenschaften, D-85748~Garching, Germany}
\affiliation{Physik-Department, Technische Universit\"{a}t M\"{u}nchen, D-85748 Garching, Germany}

\author{A.~Marx}
\affiliation{Walther-Mei{\ss}ner-Institut, Bayerische Akademie der Wissenschaften, D-85748~Garching, Germany}

\author{E.~Solano}
\affiliation{Departamento de Qu\'{\i}mica F\'{\i}sica, Universidad del Pa\'{\i}s Vasco - Euskal Herriko Unibertsitatea, Apdo. 644, 48080 Bilbao, Spain}
\affiliation{IKERBASQUE, Basque Foundation for Science, 48011 Bilbao, Spain}

\author{R.~Gross}
\affiliation{Walther-Mei{\ss}ner-Institut, Bayerische Akademie der Wissenschaften, D-85748~Garching, Germany}
\affiliation{Physik-Department, Technische Universit\"{a}t M\"{u}nchen, D-85748 Garching, Germany}

\begin{abstract}
Quantum state reconstruction involves measurement devices that are usually
described by idealized models, but not known in full detail in experiments. For
weak propagating microwaves, the detection process requires linear
amplifiers which obscure the signal with random noise. Here, we introduce a theory which nevertheless allows one to use these devices for measuring all quadrature moments
of propagating quantum microwaves based on cross-correlations from a dual-path
amplification setup. Simultaneously, the detector noise properties are determined, allowing for tomography. We demonstrate the feasibility of our novel concept by proof-of-principle experiments with classical mixtures of weak coherent microwaves.
\end{abstract}

\date{\today}

\maketitle

The reconstruction of the Wigner function~\cite{Leonhardt:1997a} or density
matrix of a propagating quantum field represents a cornerstone in quantum
optical measurement theory and experiments. In quantum homodyne
tomography~\cite{Leonhardt:1997a,Lvovsky:2009a}, for example, the signal is
combined with a local oscillator in a beam splitter, and the intensities at the
output ports are subtracted to produce the measurement of the amplified field
quadratures in terms of a histogram. The latter gives access to all quadrature
moments, or, equivalently, the Wigner function~\cite{Leonhardt:1997a}. In this
procedure, it is of utmost importance that field amplifiers are not needed in
the optical domain because photodetectors with sufficient efficiency are
available at the single-photon level~\cite{Leonhardt:1997a}. In contrast, in
the $1{-}10\,\giga\hertz$ range, which has become highly relevant due to the
advent of circuit quantum electrodynamics (QED)~\cite{Wallraff:2004a,
Blais:2004a, Houck:2007a, Deppe:2008a, Mariantoni:2008a, Fink:2009a, Hofheinz:2009a, Niemczyk:2009a, Astafiev:2010a}, only theoretical proposals exist
for the detection of propagating single microwave photons~\cite{Helmer:2009a,
Romero:2009a, Romero:2009b}. Consequently, the detection of few-photon
microwave signals requires linear amplification. Within well-established
``off-the-shelf" technology, cryogenic high electron mobility transistor (HEMT)
amplifiers lend themselves to this purpose. They offer flat gain over a broad
frequency range, but they obscure the signals by adding random
noise~\cite{Caves:1982a,Clerk:2009a} of 10--20 photons at $5\,\giga\hertz$.
Nevertheless, we prove here that it is still possible to measure all quadrature
moments of few-photon propagating microwaves in this situation. Furthermore, we
show that our proposed reconstruction method also produces a measurement of all
quadrature moments of the detector noise. In this sense, moving from a single
amplification chain to a dual-path configuration constitutes a step beyond pure
state reconstruction and towards the complete calibration of the measurement
device, i.e., detector tomography~\cite{Lundeen:2009a}. We note that so far
only state reconstruction of the intra-cavity field has been demonstrated in
circuit QED~\cite{Hofheinz:2009a}. However, quantum states of propagating
microwaves themselves can be valuable in quantum information
processing~\cite{Kok:2007a} and their full reconstruction represents an
important open issue.

\begin{figure}[ht]
\centering{\includegraphics[width=8.5cm]{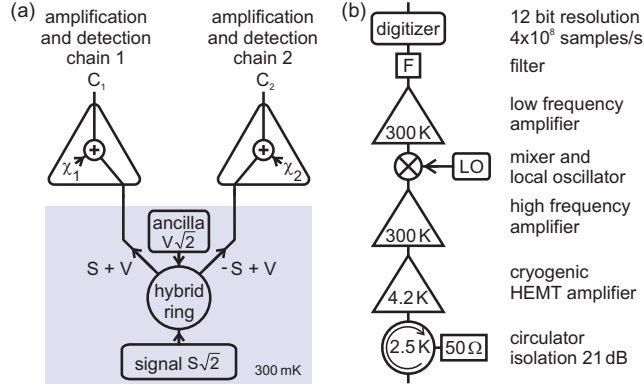}} 
\caption{
(a) Schematics of the dual path amplification and detection setup. The ancilla
port of the $180\degree$ hybrid ring is terminated by a $50\,\ohm$ load at
$T=300$\,mK. The split signals are amplified and detected in separate chains,
which add the noise $\chi_1$ and $\chi_2$ to the signal. (b) Amplification and
detection chain. The noise temperature is dominated by the cryogenic amplifiers anchored at $4.2\,\kelvin$. The isolation of the hybrid ring ($40\,\deci\bel$) and the circulator prevents correlated noise contributions from the two chains. The phase difference between $C_1$ and $C_2$ is close to $180\degree$.
}
 \label{fig:Setup}
\end{figure}

We first develop a theory for the measurement of all moments of both a
propagating quantum microwave signal and the noise added by the detector. The
basic idea is illustrated in Fig.~\ref{fig:Setup}a. A signal $S\sqrt{2}$ is
equally split at low temperatures by means of a four-port 50-50 microwave beam
splitter. The outputs  are amplified and synchronously digitized. During this
process, the amplifiers add the independent noise contributions $\chi_1$ and
$\chi_2$ to the split signals in the detection chains 1 and 2, respectively.
Assuming a $180\degree$ hybrid ring~\cite{Collin:2001a} as a beam splitter
(cf.\ appendix), the recorded time traces are   $C_1\,{=}\,G(S{+}V{+}\chi_1)$
and $C_2\,{=}\,G({-}S{+}V{+}\chi_2)$, where $G$ is the gain and $V$ represents
the ancilla state present in the fourth port of the hybrid ring. Since we
assume full knowledge of $V$, an obvious choice is the vacuum or, as in this
work, a weak thermal state. Vacuum or thermal states at gigahertz frequencies
can be prepared reliably in an experiment by controlling the
temperature~\cite{Gabelli:2004a, Mariantoni:2010a}. In the next step, we
compute suitable correlations of the form $\tav{C_1^\ell C_2^m}$, where
$\ell,m{\in}\mathbb{N}_0$ and the brackets denote ensemble averaging. In
contrast to Hanbury Brown and Twiss experiments based on intensity (power)
correlations~\cite{Gabelli:2004a}, our method is devoted to the correlations of
field quadratures (voltages). For the first signal moment, the mean values
$\tav{\chi_1}$ and $\tav{\chi_2}$ vanish and one obtains
$\tav{S}\,{=}\,\tav{C_1}/G\,{=}\,{-}\tav{C_2}/G$. All higher moments of signal
and noise can now be calculated by induction:
\begin{align}
\tav{S^n}=&-\tav{C_1^{n-1} C_2}/G^n\nonumber \\
&-\sum_{k=1}^{n-1}\sum_{j=0}^{k}{\binom{n{-}1}{k}\binom{k}{j} \tav{S^{n-k}}\tav{V^j}\tav{\chi_1^{k-j}}} \nonumber \\
&+\sum_{k=0}^{n-1}\sum_{j=0}^{k}\binom{n{-}1}{k}{\binom{k}{j} \tav{S^{n-k-1}}\tav{V^{j+1}}\tav{\chi_1^{k-j}}},
 \label{eq:Sn}\displaybreak[0]\\
\tav{\chi_1^n}=&+\tav{C_1^{n}}/G^n\nonumber\\
&-\sum_{k=1}^{n}\sum_{j=0}^{k}{\binom{n}{k}\binom{k}{j} \tav{\chi_1^{n-k}} \tav{S^{k-j}} \tav{V^j}},
 \label{eq:Xn1}\displaybreak[0]\\
\tav{\chi_2^n}=&+\tav{C_2^{n}}/G^n\nonumber\\
&-\sum_{k=1}^{n}\sum_{j=0}^{k}{\binom{n}{k}\binom{k}{j} (-1)^{k-j} \tav{\chi_2^{n-k}} \tav{S^{k-j}}} \tav{V^j}.
 \label{eq:Xn2}
\end{align}%
We note that our conjectures of equal gain in both chains and a perfect
$180\degree$ hybrid ring are not a restriction in practice. In the derivation
of the above formulas, the mutual statistical independence of $S$, $V$,
$\chi_1$ and $\chi_2$ is crucial because it implies
$\tav{S^{\beta}V^{\gamma}\chi_1^{\delta}\chi_2^{\epsilon}}\,{=}\,\tav{S^{\beta}}\tav{V^{\gamma}}\tav{\chi_1^{\delta}}\tav{\chi_2^{\epsilon}}$
for $\beta, \gamma, \delta, \epsilon {\in} \mathbb{N}_0$. The latter formula
also shows that Eqs.~(\ref{eq:Sn})-(\ref{eq:Xn2}) are suitable for quantum
signals, where $S$, $V$, $\chi_1$, and $\chi_2$ have to be interpreted as
operators.

Explicit expressions for moments of signal and detector noise can be calculated in
the spirit of Eqs.~(\ref{eq:Sn})--(\ref{eq:Xn2}). We assume
$\tav{V^{2j+1}}\,{=}\,0$ ($j\,{\in}\,\mathbb{N}$) for the ancilla as, e.g., in
the case of Gaussian statistics. We note that for the initial correlation
$\tav{C_1^\ell C_2^m}$ other choices than $\ell\,{=}\,n{-}1$ and $m\,{=}\,1$
are possible, as long as $\ell{+}m\,{=}\,n$ and $\ell,m\,{\in}\,\mathbb{N}$.
Typically, balanced products with $\ell{\approx}m$ result in simpler
expressions because of higher symmetry. Starting from $\tav{C_1C_2}$,
$\tav{C_1^2C_2}$, $\tav{C_1^2C_2^2}$ we obtain:
\begin{align*}
G^{}\tav{S_{~}^{~}}={}&\tav{C_1}=-\tav{C_2} \displaybreak[0]\\
G^2\tav{S_{}^{2}}={}&-\tav{C_1C_2}+G^2\tav{V^2} \displaybreak[0]\\
G^3\tav{S_{}^{3}}={}&-\tav{C_1^{2}C_2}-\tav{C_1}\left(\tav{C_1^2}+\tav{C_1C_2}-3G^2\tav{V^{2}}\right) \displaybreak[0]\\
G^4\tav{S_{}^{4}}={}&-G^4\tav{V^{4}}-6G^2\tav{V^{2}}\tav{C_1C_2}+6G^4\tav{V^{2}}^{2}\nonumber \\
 &+\tav{C_1C_2}^{2}+\tav{C_1^{2}C_2^{2}} -\tav{C_1^{2}}\tav{C_2^{2}} \displaybreak[0]\\
\nonumber\\
\tav{\chi_1^{}}\equiv{}&0 \displaybreak[0]\\
G^2\tav{\chi_1^{2}}={}&\tav{C_1^{2}}+\tav{C_1C_2}-2G^2\tav{V^2} \displaybreak[0]\\
G^3\tav{\chi_1^{3}}={}&\tav{C_1^{3}}+\tav{C_1^{2}C_2}-2\tav{C_1}\left(\tav{C_1^{2}}+\tav{C_1C_2}\right)\displaybreak[0]\\
G^4\tav{\chi_1^{4}}={}&\tav{C_1^{4}}-12G^2\tav{V^{2}}\tav{C_1C_2}-12G^2\tav{V^{2}}\tav{C_1^{2}}\nonumber\\
&{}+6\tav{C_1C_2}\tav{C_1^{2}}+5\tav{C_1C_2}^{2}+12G^4\tav{V^{2}}^{2}\nonumber \\
 & -4\tav{C_1}\tav{C_1^{3}}-4\tav{C_1}\tav{C_1^{2}C_2}+ 8\tav{C_1}^{2}\tav{C_1^{2}}\nonumber\\
 &+8\tav{C_1}^{2}\tav{C_1C_2}-\tav{C_1^{2}C_2^{2}}+\tav{C_1^{2}}\tav{C_2^{2}}
\end{align*}
Similar formulas can be derived for $\chi_2$. Furthermore, we assume the first
moment of the detector noise to vanish for both chains throughout the
experimental part of this work. Practically, this implies an offset correction (c.f. appendix).
Finally, from Eqs.~(\ref{eq:Sn})--(\ref{eq:Xn2}), the central moments can be
retrieved with the binomial transformation
\begin{equation}
  \tcm{S}{n}=\sum_{k=0}^{n}{\binom{n}{k}(-1)^{n-k}\tav{S^k}}\tav{S}^{n-k}\,.
  \label{eq:SCn}
\end{equation}

\begin{figure*}[t]
 \centering{\includegraphics[width=0.96\textwidth]{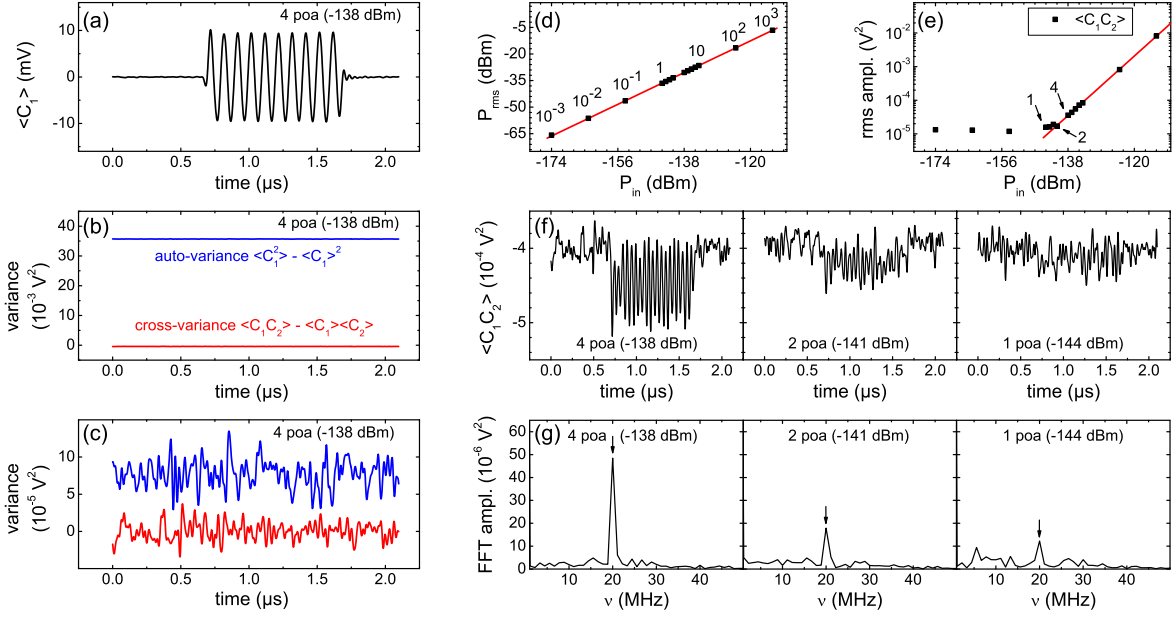}}
\caption{ Detection of coherent microwave probing pulses ($10^7$~traces
averaged): (a) Mean value of down converted signal ($10\,\mega\hertz$). (b)
Auto- (blue) and cross-variance (red). (c) Fluctuations of the  auto- and
cross-variance. The traces are obtained by subtracting the respective time
averages, and the auto-variance trace has been shifted by $8{\times}10^{-5}\,\volt^2$,
for clarity. (d) Dynamic range of the mean value. $P_\mathrm{rms}$ is the root
mean square (rms) power at the digitizer. The numbers above the data points
indicate the number of photons on average. (e) Dynamic range of the
cross-product $\tav{C_1C_2}$. The arrows indicate the values obtained from the
time traces shown in (f). The rms amplitude inside the pulse region is
calculated after subtracting the time average from each data point. (f) Power
dependence of the cross-product time traces in the limit of few photons on
average. (g) Discrete Fourier transform of the pulse region of the traces shown
in (f). The arrows indicate the $20\,\mega\hertz$ peaks. }
 \label{fig:AutoCrossRes}
\end{figure*}

We emphasize the practical relevance of the above theory by conducting
proof-of-principle experiments with weak classical microwaves. The setup is shown in Fig.~\ref{fig:Setup}.   
 As signals, we use pulsed coherent
microwaves with a frequency of $5.85\,\giga\hertz$ generated by a source at
room temperature. A series of cold attenuators ensures that the thermal noise
at the signal port of the hybrid ring is restricted to that of an effective
$50\,\ohm$-termination at the base temperature of $300\,\milli\kelvin$. The
source power at the input of the hybrid ring, $P_\text{in}$, is related to an
equivalent number of signal photons on average (poa) as described in the appendix.
Figure~\ref{fig:AutoCrossRes}a shows the ensemble average of a typical signal
used in our experiments. The pulse duration of 1$\,\micro\second$ mimics
standard cavity decay times in circuit QED
experiments~\cite{Niemczyk:2009a,Goeppl:2008a}. We first demonstrate the
suppression of the amplifier noise via cross-correlations. The auto-variance
$\tav{C_1^2}{-}\tav{C_1}^2$ of the ensemble is depicted in
Fig.~\ref{fig:AutoCrossRes}b, where one immediately notices the large offset of
$35.7{{\times}}10^{-3}\,\volt^2$ due to the amplifier noise. In the cross-variance
$\tav{C_1 C_2}{-}\tav{C_1}\tav{C_2}$, this offset is efficiently suppressed by
two orders of magnitude. As expected for a coherent signal, the variances are
flat and do not allow to distinguish between the ``on''- and ``off''-regions of
the pulses. The fluctuations of the variance signals are smaller for the
cross-correlation than for the auto-correlation by a factor of 1.6, see
Fig.~\ref{fig:AutoCrossRes}c. Next, we prove that our method works efficiently
at the quantum level, i.e., for signals of few photons on average. To this end,
we investigate the resolution limits of the constituents of the variance, mean
value and cross-product. In Fig.~\ref{fig:AutoCrossRes}d, the root mean square
power inside the pulse region is plotted against the signal power at the input
of the hybrid ring. We find a large dynamic range of the mean value extending
over six decades down to 0.001\,poa. This means that pulse energies as low as
$3.7 {\times} 10^{-26}\,\joule$ (0.01\,poa) are still very well detectable. The power
dependence of the cross-product is displayed in
Figs.~\ref{fig:AutoCrossRes}e--g. Down to 2\,poa, the pulse region is clearly
visible (Fig.~\ref{fig:AutoCrossRes}f). For 1\,poa, a Fourier transform
(Fig.~\ref{fig:AutoCrossRes}g) reveals that the signal component can still be
identified. However, the associated peak has approximately the same amplitude
as the largest noise peak in the spectrum. Hence, the detection limit of the
cross-product (see also Fig.~\ref{fig:AutoCrossRes}e) and therefore the one of
the cross-variance is \mbox{1--2\,poa}. We note that this is not a fundamental
limit, but is rather due to technical issues such as the ensemble size, filter
bandwidth, or bit resolution of the digitizer card.

\begin{figure}[t]
\includegraphics[width=8.4cm]{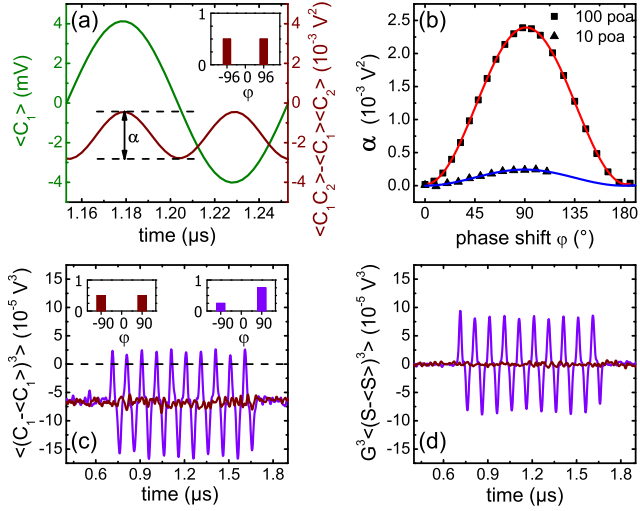}
\caption{Statistical mixtures of phase-shifted pulses: (a) Mean value and
cross-variance. The inset shows the pulse distribution for $\varphi\,{=}\,96
\degree$.  The amplitude $\alpha$ is extracted with a numerical fit. (b)
Cross-variance amplitude $\alpha$ plotted versus $\varphi$. Squares, 100\,poa.
Triangles, 10\,poa. The solid lines are fits to the data. (c) Third central
moment at 100\,poa measured with a single amplification chain. Note that, as
expected, the trace is not a simple sine function, but $\propto \sin^3(\omega
t)$. The insets show the histograms: equally distributed (brown) and with
finite skew (violet). (d) Third central moment measured with the dual path
method. The dataset is the same as in (c). }
 \label{fig:NTV}
\end{figure}

So far, we have studied time-independent ensemble variances because $\langle
f(t)^2 \rangle {-} \langle f(t) \rangle^2 \,{=}\,0$ always holds for
deterministic signals. Time-dependent variance signals require statistical
mixtures of signals. As a first example, a sequence of coherent microwave
pulses with alternating phase shifts $\pm\varphi$ is applied to the input line.
This corresponds to a statistical mixture with an equally distributed
histogram. The mean value and cross-variance are $A\cos(\varphi)\sin(\omega t)$
and $-\alpha \cos^2(\omega t)$, respectively, where
$\alpha\,{=}\,A^2\sin^2(\varphi)$ is the cross-variance amplitude and
$\omega/2\pi\,{=}\,10\,\mega\hertz$ is the signal frequency after the mixers.
The frequency doubling $\cos^2(\omega t) {\propto} [1{+}\cos(2\omega t)]$ is
confirmed by the data shown in Fig.~\ref{fig:NTV}a. Repeating the experiment
for different phase shifts, we reproduce the expected
$\sin^2(\varphi)$-dependence of $\alpha$ for 10 and 100\,poa (cf.\
Fig.~\ref{fig:NTV}b).

Studying signal mixtures also allows us to address time-dependent third central
moments $\tav{(S{-}\tav{S})^3}$ and $\tav{(\chi_1{-}\tav{\chi_1})^3}$. They are
proportional to the skewness of their associated probability histograms and
vanish for Gaussian distributions. In particular, both quantities are zero for
an equally distributed histogram. Hence, to  observe an oscillating third
central moment a statistical mixture with an asymmetric histogram (cf.\ inset
of Fig.~\ref{fig:NTV}c) is required. In the experiment, we again create a train
of pulses with 100\,poa, but this time 75\% of them are shifted by the phase
$\varphi$ and 25\% by $-\varphi$. Figure~\ref{fig:NTV}c shows the third
central moment calculated from the data of a single amplification chain. In
contrast to the case of an equally distributed mixture, a clear oscillating
signal is visible within the pulse duration for a skewed histogram. However,
there is an offset of approximately $-7\,{\times}10^{-5}\,\volt^3$. In
Fig.~\ref{fig:NTV}d, one can see that this offset disappears when also taking
into account the data from the second chain and applying the dual-path
detection scheme described by Eqs.~(\ref{eq:Sn})-(\ref{eq:SCn}). Following
these equations, we can write down the third central moment of $C_1$ as the sum
of the third central moments of signal, noise, and ancilla state:
\begin{align}
\tav{(C_1-\tav{C_1})^3}={}&G^3\tav{(S-\tav{S})^3}+G^3\tav{(\chi_1-\tav{\chi_1})^3}
 \nonumber\\
 &{}+G^3\tav{(V-\tav{V})^3}.\nonumber
\end{align}
Since $\tav{(V{-}\tav{V})^3}\,{=}\,0$ due to the the Gaussian statistics of
$V$, it becomes obvious that the offset in Fig.~\ref{fig:NTV}c must be
$G^3\tav{(\chi_1{-}\tav{\chi_1})^3}$. In this sense, the noise of the detection
chain shows non-Gaussian statistics. Although the exact origin of the latter
inside the detection chain remains unclear, our experiments provide a first
confirmation that the dual-path method is indeed capable of characterizing
signal and detector noise moments simultaneously.

In conclusion, we present a method based on off-the-shelf technology and signal
recovery techniques to gain access to arbitrary moments of both weak
propagating quantum microwaves and the detector noise. In particular, we allow
for the use of linear amplifiers adding 10--20 noise photons to the signals. We
find that it is crucial to move from a single to two amplification and
detection chains and successfully perform proof-of-principle experiments with
statistical mixtures of weak pulsed coherent microwaves. We demonstrate
sufficient sensitivity for the first two moments and observe indications for a
non-Gaussian statistics of the detector noise. The experiments indicate that
our dual-path method is a suitable tool for detecting propagating quantum
signals such as squeezed states from a Josephson parametric
amplifier~\cite{Yamamoto:2008a,Castellanos-Beltran:2008a,
Bergeal:2009a,Zagoskin:2008}, Fock states~\cite{Hofheinz:2009a} leaking out of
an on-chip resonator~\cite{Houck:2007a} or non-classical microwave field states
generated in a two-resonator circuit QED setup~\cite{Mariantoni:2008a}.

\section*{Appendix}

\textit{The ancilla state} is assumed to be well known and, hence, its quadrature moments can not be detected by our method. This is not a restriction in practice because vacuum or thermal states at gigahertz frequencies can be prepared reliably in an experiment by controlling the temperature~\cite{Gabelli:2004a, Mariantoni:2010a}. Furthermore, it is more relevant to characterize quantum states such as Fock states, where part of the Wigner function is negative. 

\textit{Beam splitter types for the dual-path method.} Although in this work we always assume a $180\degree$ hybrid ring as beam splitter, other choices are possible, as long as they provide enough isolation between the output ports. However, one has to keep in mind that all lossless, matched and reciprocal beam splitters must be four port devices~\cite{Collin:2001a}. Sometimes, as in the case of the Wilkinson power divider~\cite{Collin:2001a,Mariantoni:2010a}, the fourth port may be hidden internally and therefore complicate the analytical treatment of the ancilla state. Nevertheless, the equivalent of Eqs.~(1)--(3) of the supplemented paper can in principle be calculated for any beam splitter.

\textit{Offset correction.} The raw data is divided into segments of 4128 traces equivalent to 0.5\,s
measurement time. For each of these segments, the time average is subtracted
from each data point before any other manipulation. Effectively, this procedure
acts as a high-pass filter eliminating slow drifts in the data.

\textit{The calibration of the input line} can be performed {\it in situ} in our setup. To this end, the $50\,\ohm$ load at the ancilla port of the hybrid ring is temperature controlled using a heater and a  thermometer. Except for small deviations, we can assume the hybrid ring, ancilla load, and effective $50\,\ohm$ load at the signal port to have the same temperature. Both loads inject thermal voltage fluctuations into the hybrid ring. The auto-variance of these fluctuations follows the well-known Planck function.
Knowing the bandwidth of our setup from measurements with a spectrum analyzer ($51\,\mega\hertz$), the gains of the amplification and detection chains are inferred from a numerical fit of the Planck function to the auto-variance data~\cite{Mariantoni:2010a}. We find gains of approximately 110\,dB for both chains, where small residual gain asymmetries have been absorbed in a compensation factor already. Together with the $0.5\,\deci\bel$ loss of the hybrid ring and the total transmission from source to digitizer, we can extract an input line attenuation of $94\,\deci\bel$. From this, we determine the signal power $P_\text{in}$ at the input of the hybrid ring. The corresponding number of photons on average is the pulse energy $P_\text{in}T_\text{pulse}$ divided by the energy quantum $h{\times}5.85\,\giga\hertz$. Here, $h$ is the Planck constant and $T_\text{pulse}\,{=}\,1\,\micro\second$ the pulse duration.

\section*{Acknowledgments}
We thank C.~Probst, K.~Neumaier and K.~Uhlig for providing their expertise in
cryogenic engineering and M.~H\"aberlein and C.~Rauh for programming support.
Furthermore, we acknowledge financial support by the Deutsche
Forschungsgemeinschaft via SFB~631, the German Excellence Initiative via NIM,
the European project EuroSQIP, UPV-EHU Grant GIU07/40 and Ministerio de Ciencia
e Innovaci\'on FIS2009-12773-C02-01.

\section*{Author contributions}
E.P.M. is responsible for the main contributions regarding the theoretical and experimental results presented in this work.
F.D. provided important contributions in theory and experiment.
E.P.M. and F.D. prepared the manuscript.
M.M. provided valuable input to the initial theoretical ideas and helped during the experiments.
E.S. supervised the theoretical aspects of this work.
M.A.A.C. and A.B. helped with the experimental setup.
T.N. and E.H. contributed to discussions and helped editing the manuscript.
A.M. and R.G. supervised the experimental aspects of the project.

\end{document}